\documentclass{osa-article}
\usepackage{amsmath} 
\usepackage{mathrsfs}
\usepackage{hyperref}
\usepackage[titletoc,title]{appendix}

\journal{oe}

\usepackage{physics}

\begin{document}

\title{Phase gradient protection of stored spatially multimode perfect optical vortex beams in a diffused rubidium vapor 
}
 
\author{Yun Chen,\authormark{1} Jinwen Wang,\authormark{1,2} Chengyuan Wang,\authormark{1,4} Shougang Zhang,\authormark{3} Mingtao Cao,\authormark{3,5} Sonja Franke-Arnold,\authormark{2} Hong Gao,\authormark{1} and Fuli Li\authormark{1}}

\address{\authormark{1}Ministry of Education Key Laboratory for Nonequilibrium Synthesis and Modulation of Condensed Matter, Shaanxi Province Key Laboratory of Quantum Information and Quantum Optoelectronic Devices, School of Physics, Xi'an Jiaotong University, Xi'an 710049, China\\
\authormark{2}School of Physics and Astronomy, University of Glasgow, G12 8QQ, United Kingdom\\
\authormark{3}Key Laboratory of Time and Frequency Primary Standards, National Time Service Center, Chinese Academy of Sciences, Xi'an 710600, China\\
\authormark{4}wcy199202@gmail.com\\
\authormark{5}caomingtao1987@163.com
}




\begin{abstract}
We experimentally investigate the optical storage of perfect optical vortex (POV) and spatially multimode perfect optical vortex (MPOV) beams via  electromagnetically induced transparency (EIT) in a hot vapor cell. In particular, we study the role that phase gradients and phase singularities play in reducing the blurring of the retrieved images due to atomic diffusion. Three kinds of manifestations are enumerated to demonstrate such effect. 
Firstly, the suppression of the ring width broadening is more prominent for POVs with larger orbital angular momentum (OAM). Secondly, the retrieved double-ring MPOV beams' profiles present regular dark singularity distributions that are related to their vortex charge difference. Thirdly, the storage fidelities of the triple-ring MPOVs are substantially improved by designing line phase singularities between multi-ring MPOVs with the same OAM number but $\pi$ offset phases between adjacent rings. Our experimental demonstration of MPOV storage opens new opportunities for increasing data capacity in quantum memories by spatial multiplexing, as well as the generation and manipulation of complex optical vortex arrays.
\end{abstract}

\section{Introduction}
Quantum memories play a cornerstone role in building a global-scale quantum internet \cite{wehner2018quantum} and have attracted considerable research interests over the past few decades \cite{RevModPhys.77.633,PhysRevLett.86.783,guo2019high,saglamyurek2018coherent}. Photons encoded in different degrees of freedom (DOFs), e.g., polarization \cite{Cho:10,wang2019efficient}, frequency \cite{seri2019quantum}, time-bin \cite{tang2015storage} and spatial mode \cite{ding2013single,Wang:19,Ye:19,Yu:21,wang2020vectorial}, could be effectively stored in coherent media. Orbital angular momentum (OAM) is associated with a representative spatial DOF and has many potential applications in quantum information technologies. Laguerre–Gaussian (LG) beams are perhaps the most well-known vortex beams, carrying an OAM of $l\hbar$ per photon, where the integer $l$ denotes their charge, and various experiments have demonstrated the storage of such beams\cite{PhysRevLett.98.203601,nicolas2014quantum,PhysRevLett.114.050502,ding2019broad,wang2020efficient}. One drawback of LG modes is that their beam waists expand with increasing OAM values. As a result, for a coherent medium with finite cross-sectional dimensions, the memory efficiency for high-order LG modes will undergo tremendous decline\cite{PhysRevLett.114.050502,ding2019broad}. A perfect optical vortex (POV) beam can also carry OAM,
but its intensity distribution and radial dimension is independent of the OAM value
\cite{ostrovsky2013generation,Vaity:15}. Considering a more profound perspective for further increasing the information capacity, similar to spatial-mode multiplexing \cite{PhysRevA.86.023801}, one can nest multiple POVs with independent OAM to form a spatial multimode POV (MPOV)\cite{7893689}, with broad application prospects in high-capacity information processing. To date, research devoted to POVs and MPOVs mainly focuses on generation \cite{pu2015catenary,liu2017generation,anaya2017generation,liu2021broadband}, property analysis \cite{ma2017situ,Pinnell:19,PinnellJ:19,chu2020hybrid}, optical manipulation \cite{chen2013dynamics,garcia2014simple,tkachenko2017possible} and optical communication \cite{shao2018free,karahroudi2018performance,li2018perfect,yang2020beam,brunet2014design}. A few studies also explore quantum applications and the nonlinear effects based on POVs \cite{jabir2016generation,pinnell2020experimental}. 
But there is still no work on investigating the storage performance of POV and MPOV beams.

In this paper, we provide the first investigation on the storage of POV and MPOV beams in a warm atomic vapor cell. We begin by investigating the storage performance of POV beams with different OAM. Our observations show that the retrieved POV beams' ring widths continuously expand with increasing storage time, which mainly results from the atomic thermal motion-induced diffusion. We then generalize our experiment to the storage of double-ring MPOV beams containing two nested POV beams with different OAMs.  We find that for small ring spacings there is cross-talk between the individual POV beam constituents, except at azimuthal angles where there is a $\pi$ phase difference. 
Moreover, we investigate the role of phase patterns in specifically phase singularities for the storage of more complex triple-ring MPOVs. The results manifest that the phase distributions containing a singularity line which separates the individual rings perform better in resisting image distortion caused by atomic diffusion and enable higher storage fidelity.

\section{Theoretical description}
Generally, a POV beam can be understood as the Fourier transformation of a Bessel beam \cite{Vaity:15}, whose complex field amplitude can be expressed in polar coordinates ($r, \phi$) as 
\begin{equation}\label{Eq:1}
E_{\rm{POV}}=A_{0}\;{\rm{exp}}[-(r-R)^{2}/w^{2}]\;\exp [-i (l \phi+\Phi_{0})],
\end{equation}
where $A_{0}$, $\Phi_{0}$, and $l$ are the amplitude, initial constant phase, and vortex charge of the POV, respectively, while $R$ and $w$ represent the radius and half width of the POV ring.

MPOVs can be obtained as linear sums of multiple such POV beams, with $R_{m}$, $l_{m}$ and $\Phi_{m}$ denoting the radius, OAM and initial phase of the $m^{th}$ sub-POV.  For simplicity we keep $A_0$ and $w$ constant for all constituent sub-POVs, 
obtaining the field amplitude of the MPOV as:
\begin{equation}\label{Eq:2}
E_{\rm{MPOV}}=A_0 \sum_{m}{\rm{exp}}[-(r-R_{m})^{2}/w^{2}]\;{\rm{exp}}[-i(l_{m}\phi+\Phi_{m})],
\end{equation}
where we use the convention of labeling the rings from the innermost to the outermost.

\begin{figure}[ht!]
\centering\includegraphics[width=9cm]{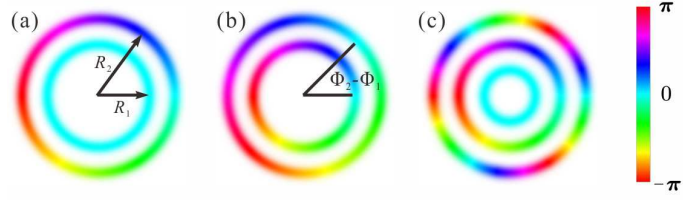}
\caption{Simulated intensity and phase structure (shown as hue color) of representative MPOVs. (a) Double-ring MPOV with OAMs $l_1=0,\; l_2=3$ at radii $R_1=6 w, \; R_2=9 w$, and equal phase offset. (b) Double-ring MPOV with $l_1=l_2=1$ at radii $R_1=6 w, \; R_2=9 w$, and phases $\Phi_1=0,\; \Phi_2=\pi/4$. (c) Triple-ring MPOV with $l_1=0,\; l_2=1,\; l_3=3$ at radii $R_1=3 w, \; R_2=6 w,\; R_3=9 w$, and no phase offset. }
\label{fig:1}
\end{figure}

In our following experiments, we set $R_{m+1}-R_m > 2w$, guaranteeing the ring spacing between two adjacent sub-POVs large enough to avoid the formation of visible phase singularities in their overlap region, as mentioned in \cite{Li:18}. As seen in Fig.~\ref{fig:1}, the POVs form a concentric ring structure and each ring carries an adjustable OAM. Similar to general LG beams, which use the radial as well the azimuthal degree of freedom to store information, these MPOVs allow utilize the radial degree of freedom for increasing information capacity, however with less restrictions on the radial and azimuthal mode structure.

In an EIT quantum memory, the MPOV probe in conjunction with a control beam drives a $\Lambda$ transition between atomic hyperfine levels, thereby mapping the MPOV field onto a stationary dark state of the atomic ground state coherence. 
However, atomic thermal motion-induced diffusion is inevitable, rendering the beam's pattern stored in one location be released elsewhere. As a result, the retrieved image becomes blurred and its visibility declines. M. Shuker \emph {et al.} studied this diffusion process both theoretically and experimentally \cite{PhysRevLett.86.783}. Based on their theoretical model, the retrieved pattern of the MPOV after stored from $t_{0}$ until $t_{0}+t$ is given by
\begin{equation}\label{Eq:3}
E_{\rm{MPOV}}(\textbf{r}, t_{0}+t) \propto \int d^{2}{\tilde{\textbf{r}}}E_{\rm{MPOV}}(\tilde{\textbf{r}}, t_{0})G(\textbf{r}-\tilde{\textbf{r}}),
\end{equation}
where $G(\textbf{r})$ is the diffusion propagator which can be expressed as $G(\textbf{r})=\exp[-\abs{r}^2/(4 D t_{0})]$, $D$ is the atomic diffusion coefficient, $E_{\rm{MPOV}}(\tilde{\textbf{r}}, t_{0})$ is the field to be stored. According to the convolution theorem, one can rewrite Eq. (\ref{Eq:3}) as 
\begin{equation}\label{Eq:4}
E_{\rm{MPOV}}(\textbf{r}, t_{0}+t) \propto \mathcal{I F}[\mathcal{F} [{E_{\rm{MPOV}}({\textbf{r}}, t_{0})}] \mathcal {F} [{G(\textbf{r})}]],
\end{equation}
where $\mathcal{F}$ and $\mathcal{IF}$ represent a Fourier transform (FT) and inverse Fourier transform (IFT), respectively. 

\section{Experimental setup}
\begin{figure}[bp]
\centering\includegraphics[width=12cm]{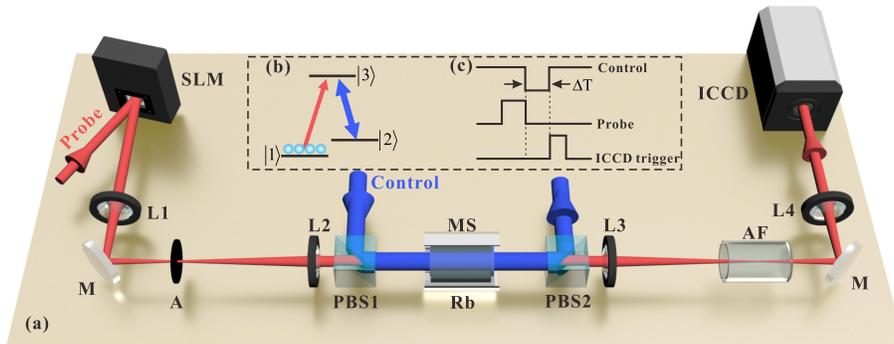}
\caption{(a) Experimental setup. SLM: spatial light modulator, L1-L4: lenses with a focal length of $f=500$ mm, M: mirror, A: aperture, PBS1, PBS2: polarizing beam splitters, MS: magnetically shielded oven, Rb: 5 cm long $^{87}$Rb atomic vapor cell at 60 $^\circ$C, AF: atomic filter, ICCD: intensified charge-coupled device camera. The powers of the control beam and the probe beam are 18 mW and 100 $\mu$W, respectively. (b) Relevant energy levels used in the experiment. (c) Time sequence for optical storage.}\label{fig:2}
\end{figure}
The experimental setup for generating and storing POV and MPOV beams is illustrated in Fig.~\ref{fig:2}(a). A horizontally polarized probe is modulated to a POV (MPOV) beam via a spatial light modulator (SLM) with the method in Ref. \cite{Bolduc:13}. Several representative MPOVs generated in our experiment and their phase pattern characterizaton are given in Appendix \ref{Appendix A}. The modulated probe is then imaged to the center of the atomic vapor cell by the first 4$f$ imaging system (composed of L1 and L2). The vapor cell containing 8 Torr neon buffer gas is heated to 60$^{\circ}$C, rendering a diffusion coefficient of $D \approx 25$ cm$^2$/s (more details are shown in Appendix \ref{Appendix B}). A vertically polarized control beam is superimposed with the probe by the polarizing beam splitter 1 (PBS1) and co-propagates with the probe beam through the atomic sample.
The control beam's waist is 6 mm, completely covering the probe beam. After the atomic cell, PBS2 and an atomic filter (AF) filter out the control beam. The retrieval probe beam is recorded by an intensified charge-coupled device camera (ICCD) after the second 4$f$ system (composed of L3 and L4). 

We adopt the $\emph{D}_{1}$-line transition of  $^{87}$Rb as depicted in Fig.~\ref{fig:2}(b), where $\ket{1}$, $\ket{2}$, and $\ket{3}$ represent $\left|5 \mathrm{S}_{1 / 2}, F=1\right\rangle$, $\left|5 \mathrm{S}_{1 / 2}, F=2\right\rangle$ and $\left|5 \mathrm{P}_{1 / 2}, \mathrm{F}^{\prime}=1\right\rangle$, respectively. Both transitions are set resonant to satisfy the EIT condition. The probe (795 nm, Toptica DL pro) and control lasers (795 nm, Toptica DL) are phase-locked by an offset phase lock servo (Vescent D2-135), and they are chopped separately by two acousto-optic modulators (AOMs), allowing us to precisely control the time sequence as shown in Fig.~\ref{fig:2}(c). The probe beam's pulse width is 2 $\mu$s, and $\Delta {\rm{T}}$ denotes the storage time. The trigger of the ICCD synchronizes with the retrieval process.

\begin{figure}[tp]
\centering\includegraphics[width=10cm]{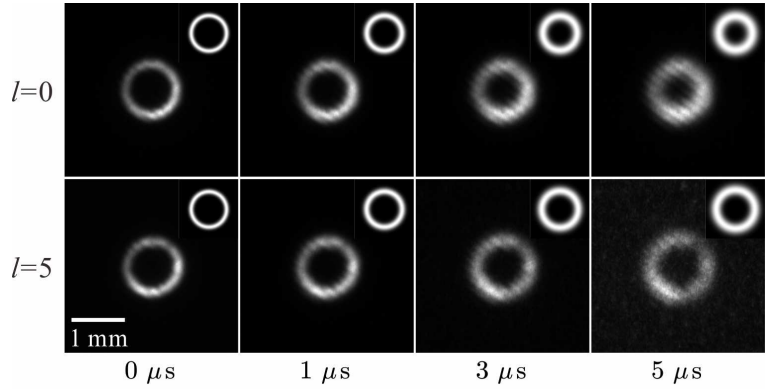}
\caption{Intensity profiles of retrieved POV beams with $l=0$ (a(1)-a(4)) and $l=5$ (b(1)-b(4)) versus various storage time ($\Delta {\rm {T}}=1, 3, 5$ $\mu$s). $\Delta {\rm{T}}=0$ $\mu$s indicates the unstored probe pattern. The insets are simulations.}\label{fig:3}
\end{figure}
\section{Experiment results and discussions}
\subsection{Storage of single-ring POV}
\begin{figure}[bp]
\centering\includegraphics[width=11cm]{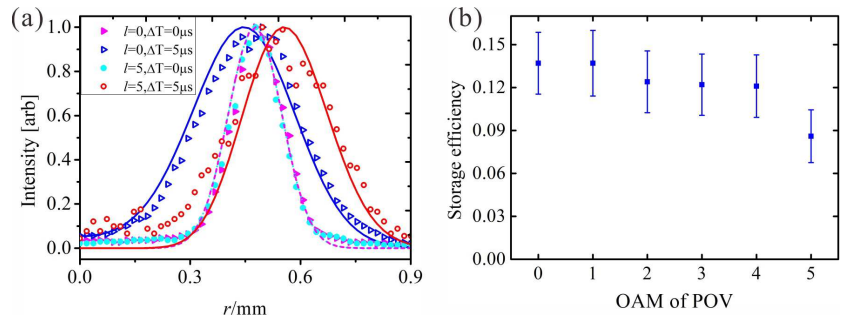}
\caption{(a) The intensity distributions along the radial direction from the center of the retrieved POVs. The dashed and solid lines correspond to the theoretical curves at $\Delta$T = 0 $\mu$s and $\Delta$T = 5 $\mu$s, respectively. (b) The storage efficiency versus the OAM of the POV probe for $2\;\mu$s storage. Error bars represent one standard deviation due to fluctuations of measured storage efficiencies.}\label{fig:4}
\end{figure}
We firstly investigate the storage of general single-ring POVs, for the examples of $l=0$ and $l=5$. The adpoted POVs have a radius of 0.45 mm and a half width-ring of 0.14 mm. Figure~\ref{fig:3} shows the retrieved patterns of the POVs at incremental storage durations ($\Delta$T = 0, 1, 3, 5 $\mu$s). The insets (at the top right of all the figures in this paper) are the simulation results based on Eq.~(\ref{Eq:4}). As the storage time increases, the ring widths of the retrieved POVs get notably broader, which can be explained largely from the atomic thermal motion-induced diffusion. A second, more subtle effect is that broadening at the inner edge of the intensity ring is more pronounced for POVs with $l=0$. We analyze this quantitatively by plotting the intensity distributions and the corresponding theoretical predictions along the radial direction from the center of the retrieved POVs, as shown in Fig.~\ref{fig:4}(a). As the same with the previous works \cite{ostrovsky2013generation,Vaity:15}, the generated intensity profiles (correspond to $\Delta$T = 0 $\mu$s) for POVs with different OAM are almost identical (cyan and magenta dashed lines), but are broadened to varying degrees after $5\;\mu$s storage time. The intensity peak of the retrieved POV with $l=0$ is shifted inward by 0.26 $w$ and its full width at half maximum (FWHM) is increased by 97\%. In comarision, the intensity peak of the retrieved POV with $l=5$ is shifted outward by 0.51 $w$ and its FWHM is increased by 80\%, while there is no obvious broadening at its inner edge.
This can be explained qualitatively by the fact that the phase gradient is larger for POVs with larger winding numbers, and larger on the inner edge ($\nabla\Phi_\phi =\frac{l}{R - w}$) than the outer edge ($\nabla\Phi_\phi =\frac{l}{R+w}$). 
Atoms diffusing through a rapidly oscillating phase pattern experience destructive interference \cite{PhysRevLett.98.203601,2010Self}, limiting diffusion for $l=5$, particularly at the inner edge. By contrast, in the case of $l = 0$, there is no destructive interference due to the plane phase, and diffusion follows Eq.~\ref{Eq:4}, filling in the beam center. It is interesting to note that for an LG mode after diffusion of duration $t$, the waist radius increases with the scaling factor $s(t)=(1+4Dt/w_{0}^{2})^{1/2}$ with $w_{0}$ the original waist, that is, the retrieved LG mode after diffusion will always expand outward. While for a retrieved POV beam, whether it contracts inwards or expands outwards depends on the phase gradient, which is determined by the POV’s parameter setting \cite{Pinnell:19}. For instance, in our experiment the retrieved POV with $l \le 2$ contracts inwards and expands outwards when $l > 2$. This unique characteristic will facilitate us to quantitatively measure and analyze the phase gradient dependent diffusion \cite{PhysRevLett.98.203601} and other optical vortices related effects such as the transverse azimuthal dephasing \cite{PhysRevA.95.033823} in the thermal atomic vapor cell.

We also provide the storage efficiency corresponding to different OAM after $2\;\mu$s storage duration (see Fig.~\ref{fig:4}(b)) by replacing the ICCD with a photodetector. Although the spot size of the POV beam does not change as the OAM number increases, the storage efficiency shows a downward trend. We attribute this to the fact of OAM related transverse azimuthal spin-wave dephasing caused by the atomic thermal motion \cite{PhysRevA.95.033823}.

\subsection{Storage of double-ring MPOV}
Next, we investigate the storage of double-ring MPOVs formed by superimposing POVs with different OAM. The wavepacket of a double-ring MPOV expresses by Eq.~(\ref{Eq:2}) with $m$ taking the values $1$ and $2$.  In our experiment we chose the radii of the rings to be $R_1=0.56\;$mm and $R_2=0.83\;$mm respectively, with $w=0.11$ mm and both initial phases set to $\Phi_1=\Phi_2=0$.
\begin{figure}[tp]
\centering\includegraphics[width=11cm]{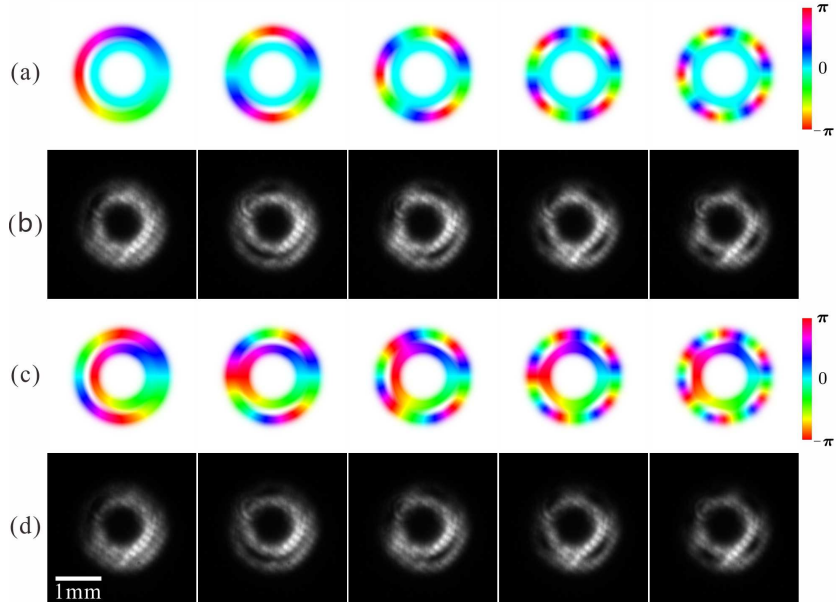}
\caption{
 Theoretical and experimental results of the retrieved double-ring MPOVs with $R_{1}$ = 0.56 mm and $R_{2}$ = 0.83 mm. (a) and (c) indicate the simulated intensity and phase structure of the retrieved fields, while (b) and (d) are their corresponding experimentally obtained intensity distributions. The storage time is fixed at 2 $\mu$s}\label{fig:5}
\end{figure}

Figure ~\ref{fig:5} shows the theoretical and experimental results of the retrieved double-ring MPOVs stored for 2 $\mu$s. For the first five realizations, we use a plane phased inner ring with $l_1=0,$ and an outer ring OAM of $1 \leq l_2 \leq 5$. The associated phase structures contain $| l_2-l_1|$ (here between one and five) singularities, positioned along a ring with radius $(R_1+R_2)/2$, at angles where the phase differences of the individual POVs are $\pm \pi$, as shown in Fig.~\ref{fig:5}(a). While the optical intensity in the area between the two rings of the sub-POVs is low due to the rapid radial Gaussian decay, it becomes true zero only at the position of the singularities due to quantum interference. Similar effects have been noted for superpositions of LG beams \cite{FrankeArnold2007} and for POV beams \cite{Li:18}.
This becomes more evident in the retrieved intensity distributions after storage, shown in Fig.~\ref{fig:5}(b).  
Due to diffusion, the two retrieved rings merge together, except at the positions of the singularities.
Similar effects occur also when both rings contain non-vanishing OAM. We investigate examples where $l_{1}$ = 1 and $l_{2}$ varies from 2 to 6. Again, between one and five singularities form, and the corresponding positions remain unfilled after storage, as can be seen from the retrieval patterns in Fig.~\ref{fig:5}(d). These results are in good agreement with simulations (Fig.~\ref{fig:5}(c)). For MPOVs with larger OAM different or longer storage time, the retrieval patterns follow similar rules except for the lower storage efficiencies.

The fact that topological structures, namely phase singularities, are protected under storage has the same physical origin as the reduced diffusion for POVs carrying larger OAM: Phase singularities are associated with strong phase gradients, leading to destructive interference and thus suppressed diffusion.

These results reveal the negative effect of atomic diffusion on the retrieved beam profile as one of the limiting factors for storing spatial multimodes in a hot atomic ensemble. To improve fidelity we could utilize two strategies: (1) The ring spacing between neighboring sub-POVs should be larger than the diffusion distance under a certain storage time. For instance, we increase the spacing between the sub-POVs and there is no diffusion-induced cross-talk. We further demonstrate that the retrieved MPOV can be de-multiplexed with each sub-POV be characterized separately. More details can be found in Appendix \ref{Appendix C}. (2) Phase gradients between neighboring sub-POVs should be optimized. In this context, it may be interesting to note that the individual rings of general LG modes are indeed separated by singularity lines, associated with a $\pi$ phase jump, and hence would be diffusion protected. 
More generally, by taking advantage of the atomic diffusion one can coherently control the evolution of the MPOV beams\cite{Shwa:12} in the vapor cell to generate versatile multi-singularity vortex beams\cite{Li:18,shen2019optical,ma2017generation,li2020anomalous,wang2021tailoring}, with potential applications in optical tweezers and optical (quantum) communication.

\begin{figure}[tp]
\centering\includegraphics[width=10cm]{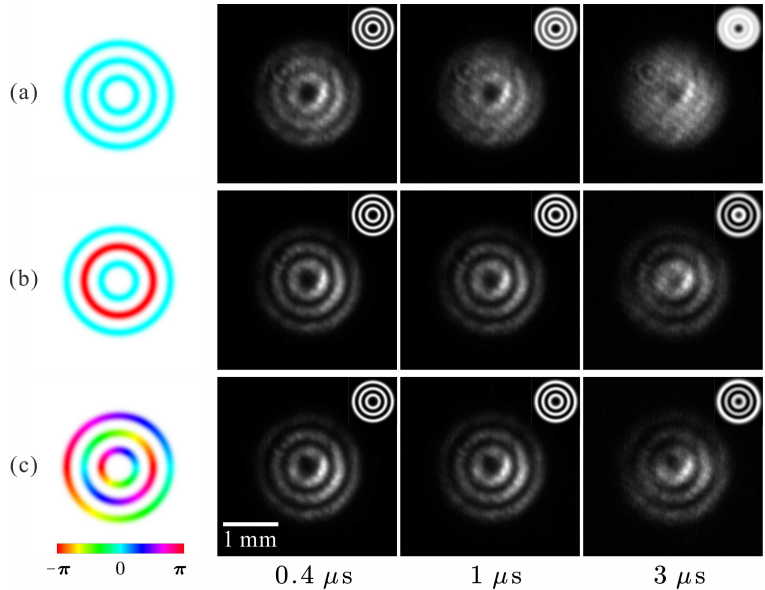}
\caption{Storing triple-ring MPOVs. The first column is the phase patterns of input MPOVs, and the second column to the fourth column are the retrieved intensity profiles versus different storage time ($\Delta {\rm {T}}$ = 0.4, 1, 3 $\mu s$). Here, $R_{1}$ = 0.28 mm, $R_{2}$ = 0.56 mm, $R_{3}$ = 0.83 mm. The insets are simulations.
}\label{fig:6}
\end{figure}
\subsection{Storage of triple-ring MPOV}
In the following, we investigate the role of phase patterns in specifically phase singularities for the storage of more complex triple-ring MPOVs, as
expresses by Eq.~(\ref{Eq:2}) with $m=1,2,3$, $w = 0.11$ mm and for radii $R_1=0.28\;$mm, $R_2=0.56\;$mm, and $R_1=0.83\;$mm respectively, see Fig.~\ref{fig:6}. 
The first column in Fig.~\ref{fig:6} shows the intensity and phase structures of adopted MPOVs, and the following columns contain the corresponding retrieved patterns for increasing storage times from $0.4$ to $3\;\mu$s. 
The MPOV in Fig.~\ref{fig:6}(a) contains flat phasefronts, with $l_{1,2,3}=0$ and $\Phi_{1,2,3}=0$; in ~\ref{fig:6}(b) the middle ring is phase shifted by $\pi$ compared to the other two rings (i.e.~$\Phi_2=\pi$); in ~\ref{fig:6}(c) all rings have $l_{1, 2, 3}=1$, with the middle ring still phase shifted by $\pi$ compared to the others. 
It is evident that phase distributions that contain a singularity line which separates the individual rings (\ref{fig:6}(b) and ~\ref{fig:6}(c)) perform better in resisting image distortion caused by atomic diffusion than those without (\ref{fig:6}(a)). This is in agreement with our interpretation of destructive interference discussed for our previous experiments, and the description in \cite{PhysRevLett.100.223601}.

Overall, when storing MPOVs, we can obtain a higher fidelity in pattern storage
by designing a suitable phase distribution. In our final section we have only investigated the storage behavior of MPOV beams with the same OAM value in all sub-POVs. For MPOVs containing different OAM values in adjacent rings a separation by a true singularity line can not be achieved, and a larger ring separation will be required. These conditions become less stringent when considering MPOV storage in other coherent media where the atoms are relatively motionless and hence diffusion is less relevant, e.g., cold atoms\cite{wang2020efficient} and crystals\cite{yang2018multiplexed}; in these cases the storage of compact multiple-ring MPOVs carrying different OAMs would be possible. 

\section{Conclusion}
In summary, we have experimentally investigated the storage and retrieval of POV and MPOV beams with the EIT scheme in a hot atomic vapor cell. We have in particular studied the role that phase gradients and phase singularities play in reducing blurring of the retrieved images due to atomic diffusion. We showed evidence for this effect in three different manifestations: Diffusion is suppressed (1) for POVs with larger OAM, and in particular at the inner edge, where the phase gradient is steeper, (2) at the positions of phase singularities embedded between multi-ring MPOVs of different OAM number, and (3) at line phase singularities between multi-ring MPOVs with the same OAM number but $\pi$ offset phases between adjacent rings. 
While we have only investigated MPOVs in this work, we expect that the same mechanism applies equally to other spatial modes.

Our experimental demonstration of MPOV storage opens new opportunities for increasing data capacity in quantum memories by spatial multiplexing.  The fact that ring radii of multi-ring MPOVs are free parameters, in combination with designing suitable phase structures may provide effective methods to improve modal fidelity during storage in thermal vapor cells. 
Moreover, the evolution of the MPOVs in the vapor cell can be coherently controlled, which may allow the generation and manipulation of more complex optical vortex arrays.

\medskip
\begin{appendices}
\section{MPOV generation and characterization}
\label{Appendix A}
\begin{figure}[bp]
\centering\includegraphics[width=10cm]{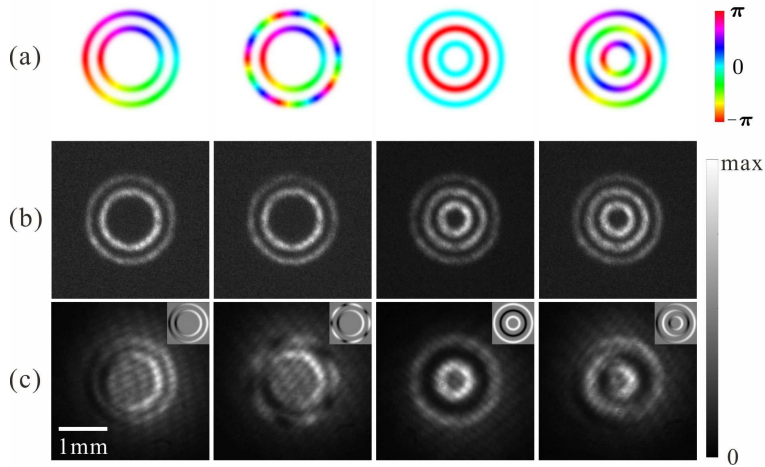}
\caption{(a) Simulated intensity and phase structure of several representative MPOVs and (b) their corresponding intensity patterns and (c) interferogram recorded experimentally. For double-ring MPOVs, $R_{1}$ = 0.56 mm and $R_{2}$ = 0.83 mm; For triple-ring MPOVs, $R_{1}$ = 0.28 mm, $R_{2}$ = 0.56 mm and $R_{3}$ = 0.83 mm. The illustrations in (c) are the corresponding theoretical predictions.}
\label{fig:7}
\end{figure}

With simultaneous intensity and phase encryption technology \cite{Bolduc:13}, various MPOVs could be generated. Figure~\ref{fig:7}(b) shows several MPOV patterns generated in our experiment, and Fig.~\ref{fig:7}(a) is their preset phase and intensity distributions. It should be mentioned that the generated intensity and the storage efficiency of the outer sub-POV are both lower than the inner sub-POV, which results in the breakage of the retrieved outer ring, as seen in  Fig.~\ref{fig:5}.

In order to verify that the generated MPOVs contain indeed the preset phase distributions, we let the MPOVs interfere with a plane wave that also originates from the probe laser, and the recorded interference patterns are exhibited in Fig.~\ref{fig:7}(c). 
We find that our experimental interferograms are in good agreement with simulated interference patterns (the insets shown in Fig.~\ref{fig:7}(c)). 

\begin{figure}[tp]
\centering\includegraphics[width=9cm]{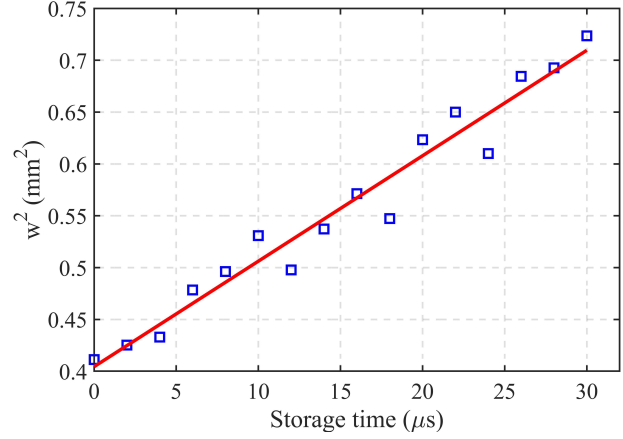}
\caption{Squared waists radii of the Gaussian beam as a function of storage time. 
}\label{fig:8}
\end{figure}

\section{Diffusion coefficient measurement}\label{Appendix B}
The diffusion coefficient $D$ is measured by storing a Gaussian beam in the hot atomic vapor. As the storage time $t$ varies from 0 to 30 $\mu$s, we record the transverse intensities of the retrieved Gaussian beams and use Gaussian fitting to obtain their waist radii $w$. The square of the waist radius versus the storage time is depicted in Fig. \ref{fig:8}. The fitting curve (red line in Fig. \ref{fig:8}) is fitted by $w(t)^{2}=w_{0}^{2}+4Dt$, where $w_{0}$ and $w(t)$ are the waist radii of the input and the retrieved Gaussian beams stored for $t$ s, respectively. From the fitting we can obtain $D\approx$ 25 cm$^{2}$/s.

\begin{figure}[t!]
\centering\includegraphics[width=8cm]{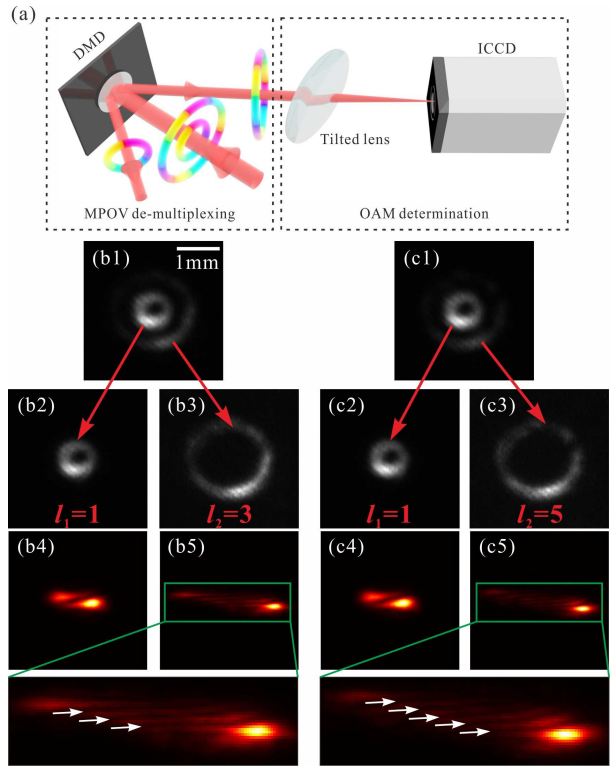}
\caption{(a) Experimental setup for MPOV de-multiplexing and OAM measurement. (b1) and (c1) are two retrieval patterns of double-ring MPOVs with $R_{1}$ = 0.28 mm and $R_{2}$ = 0.83 mm, and the corresponding sub-POVs [(b2)-(b3) and (c2)-(c3)] after de-multiplexing and their auto-interference patterns [(b4)-(b5) and (c4)-(c5)] near the focus of the tilted lens. The figures in the last row are zoom-in figures, and the white lines with arrows mark the locations of dark stripes.}\label{fig:9}
\end{figure}

\section{De-multiplexing and OAM measurement of the retrieved MPOVs}
\label{Appendix C}
As we have mentioned above, the space between two neighbouring sub-POVs should be large enough to eliminate diffusion-induced crosstalk. Figures~\ref{fig:9}(b1) and ~\ref{fig:9}(c1) show two retrieval patterns of double-ring MPOVs, $l_{1}$ = 1 \& $l_{2}$ = 3 for Fig.~\ref{fig:9}(b1) and $l_{1}$ = 1 \& $l_{2}$ = 5 for Fig.~\ref{fig:9}(c1) with larger ring spacing ($R_{1}$ = 0.28 mm, $R_{2}$ = 0.83 mm) under 2 $\mu$s storage time. There is no visible overlap between the rings. In the practical application of multiplexed light, de-multiplexing is a necessary step. Here we use an extra digital micromirror device (DMD, DLP4500) for the MPOV de-multiplexing and separate the two sub-POVs, as shown in Fig. ~\ref{fig:9}(a). The DMD is loaded with a binary amplitude mask encoded with the circular function $circ(r/r_{D})$, where $r_{D}$ satisfies $R_{1} + w <r_{D}<R_{2} - w$. The retrieved double-ring MPOV incidents perpendicularly onto the DMD, then its two de-multiplexed sub-POVs are reflected at angles of $\pm$12$^{\circ}$ with respect to the normal of the DMD. 


Besides, we also use a simple method to determine the OAM value of each de-multiplexed sub-POV. This method is based on the principle that a monochromatic beam with topological charge $l$ splits into $|l|$ tilted dark stripes in the image plane under astigmatic transformation induced by a tilted lens. The lens adopted in our experiment has a focal length of $f=50$ cm. The recorded auto-interference patterns near the focus of the tilted lens are shown in Fig.~\ref{fig:9}(b4)-\ref{fig:9}(b5) and Fig.~\ref{fig:9}(c4)-\ref{fig:9}(c5), from which we find that there are $|l|$ dark stripes across the image. These outcomes prove that the retrieved sub-POVs after de-multiplexing retain their original vortex characteristics.
\end{appendices}

\section*{Funding}
This work is supported by National Natural Science Foundation of China (NSFC) (11774286, 92050103, 12104358, 11534008, 12033007, and 61875205). 

\section*{Disclosures}
The authors declare no conflicts of interest.

\section*{Data availability}
Data underlying the results presented in this paper are not publicly available at this time but may
be obtained from the authors upon reasonable request.

\bibliography{reference}






\end{document}